\documentstyle[12pt,aasms]{article}
\journalid{337}{15 January 1989}
\articleid{11}{14}
\begin{document}

\title{TGRS OBSERVATIONS OF GAMMA-RAY 
LINES FROM NOVAE.  II. CONSTRAINING THE GALACTIC NOVA RATE FROM A SURVEY
OF THE SOUTHERN SKY DURING 1995--1997}

\author{M. J. HARRIS\footnote{Universities Space Research Association,
harris@tgrs2.gsfc.nasa.gov}, B. J. TEEGARDEN, T. L. CLINE, N. GEHRELS, 
D. M. PALMER\altaffilmark{1}, R. RAMATY, AND H. SEIFERT\altaffilmark{1}}
\affil{Code 661, NASA/Goddard Space Flight Center, Greenbelt, MD 20771}

\begin{abstract}

The good energy resolution (3--4 keV FWHM) of
the Transient Gamma Ray Spectrometer (TGRS) on board the {\em WIND\/}
spacecraft makes it sensitive to Doppler-shifted outbursts of 511
keV electron-positron annihilation radiation, the reason being that
the Doppler shift causes the cosmic line to be slightly offset from
a strong instrumental background 511 keV line at rest, which is ubiquitous
in space environments.  Such a cosmic line (blueshifted) is predicted
to arise in classical novae due to the annihilation of positrons from
$\beta$-decay on a timescale of a few hours in an
expanding envelope.  A further advantage of TGRS --- its broad field of 
view, containing the entire southern ecliptic hemisphere --- has
enabled us to make a virtually complete and unbiased 3--year search for 
classical novae at distances up to $\simeq 1$ kpc.  We present 
negative results of this search, and estimate its implications for
the highly-uncertain Galactic classical nova rate and for future space
missions.

\end{abstract}

\clearpage

\section{Introduction}

Classical novae are rather frequently observed in our Galaxy (Liller \& 
Mayer 1987, Shafter 1997), and have also been studied in external galaxies ; 
typically $N \sim$ 3--4 per year are detected in our Galaxy (Duerbeck 1995,
Warner 1995).  Most of the discoveries and observations of Galactic novae 
have been made by amateur astronomers with little access to spectroscopic 
and photometric equipment.  Sky coverage has been episodic and extremely hard 
to calculate.  Classification attempts have also been hindered.  As a result, 
many of the most basic properties
involving their global rate and distribution are surprisingly uncertain. 
For example, a number of arguments suggest that the Galactic rate of novae
must be much higher than $N$:

(a) The typical limiting apparent magnitude obtainable with amateur apparatus 
and methods has been increasing steadily in recent years, but for the 
period covered by this paper may be taken to be $m_{V} \sim 8$, within a 
very wide range, and with extremely uneven coverage.  Application of the 
expanding-photosphere method to a subset of relatively nearby and bright 
novae has yielded the empirical relation 
\begin{eqnarray}
M_{V} & = & 2.41 \log t_{2} -10.7 ~~~\mbox{for 5 d $< t_{2} < 50$ d} \\
 & = & -9 ~~~\mbox{for $t_{2} \le 5$ d} \\ 
 & = & -6.6 ~~~ \mbox {for $t_{2} > 50$ d}
\end{eqnarray}
(Warner 1995) for the absolute magnitude, where $t_{2}$ (the {\em speed
class\/}) is the time taken for $m_{V}$ to increase by 2 from
discovery.  It follows that the distance out to which amateur
astronomers are detecting typical novae is $\sim 10$ kpc, or only about one-half
the volume of the Galaxy.  Furthermore, the rate of discoveries at the
faintest magnitudes ($m_{V} > 6$) is greater than what would be 
extrapolated from brighter novae.  This indicates that a new population
--- presumably associated with the Galactic bulge rather than the disk ---
is present and poorly sampled (Duerbeck 1990; see below).

(b) Even within that part of the Galaxy which is effectively searched
for novae, the discovery rate is blatantly incomplete.  Not only does the
discovery rate for novae with $3 < m_{V} \leq 6$ fall below the 
extrapolated rate for brighter events (thus, in contrast to the
preceding argument, suggesting that many events in this range are missed:
Duerbeck 1990), but there is a marked deficiency of discoveries in the
southern celestial hemisphere (Warner 1995).  This is relevant
to our work, since the TGRS detector is permanently pointed
at the southern sky (\S 2.1).  During its period of operation
(1995--1997) five novae were discovered in the southern hemisphere
(Harris et al. 1999, hereafter Paper I), but there is no way of knowing 
how many were missed.\footnote{ The discovery process is 
particularly liable to miss
"fast" novae ($t_{2} \simeq$ a few days) which rise and fall
in between successive visits to a given location.}  The possibility of
detecting undiscovered novae as bright as
$m_{V} = 3$ (marginally within TGRS's capabilities) is one of
the justifications for the present work.

(c) In Galactic latitude, the distribution of classical novae is
somewhat concentrated toward the equatorial plane (scale heights for
disk and bulge populations 125 and 500 pc respectively: Duerbeck 1984,
1990).  They must therefore be affected to some degree by interstellar
extinction, and a deficiency of discoveries close to the plane is
indeed observed (Warner 1995).

In terms of the composition of their ejecta, novae are classified into
CO-rich and ONe-rich; it is thought that the distinction reflects the
composition of the underlying white dwarf material, with the ONe class
coming from more massive progenitors whose cores burned beyond the early
He-burning stage which yields C and O.  Different levels of positron 
annihilation line flux are expected from each class (\S 4).  If the 
progenitors of the ONe subclass are really more massive, they will tend 
to lie closer to the Galactic plane, and the resulting novae will be
more strongly affected by extinction and
relatively under-represented in the discovered sample (of which they
compose $\sim 1/3$: Gehrz et al. 1998).  Evidence of this has been
detected by Della Valle et al. (1992). 

(d) The three preceding factors would all tend to enhance the true
Galactic nova rate above that observed.  However, a second, quite distinct
approach to the problem tends to produce systematically lower rates.
In this approach, several external galaxies (particularly the Magellanic
Clouds, M31 and M33) have been monitored for novae, and their
observed rates extrapolated in some fashion to the Milky Way (Ciardullo et
al. 1987, Della Valle \& Livio 1994).
The usual basis for extrapolation is absolute blue luminosity (Della Valle
\& Claudi 1990).  As can be seen in Table 1, the results from this approach
are systematically smaller than attempts to correct for the missing
Galactic novae directly.  The original explanation for this effect was
provided by Duerbeck (1990), who postulated two different classes of
event by spatial distribution --- disk and bulge novae.  It was claimed
that the bulge population has a systematically slower speed class, and
is therefore generally less luminous by Equations (1--3), which might
account for the discrepancy, given a larger bulge in the main external
source of novae, M31.  As will be seen (\S 4.1), our search method is
probably relevant only to a disk population. 

A third approach to the problem is theoretically possible, by which
classical nova outbursts are assumed to be part of a life-cycle of
which other cataclysmic variables are manifestations.  The Galactic
nova rate is then derived from the assumed space densities of these
related objects, together with some model for the outburst
recurrence time (Warner 1995).  This approach is more reliable at 
predicting the Galactic space density rather than the global rate,
which is more directly related to the measurements we shall present.

It is important to correct for and combine these various factors into an
overall global Galactic nova rate, which would govern the input of
novae into Galactic chemical evolution, dust grains and
interstellar radioactivity (Gehrz et al. 1998).  However attempts to
do so have yielded wildly discordant 
results, ranging from 11--260 novae yr$^{-1}$ (see Table 1).

We have therefore adopted in this work yet a fourth (and simplest)
approach which is to make an {\em unbiased\/} search for novae in our Galaxy.
The detection of $\gamma$-ray lines from radioactive
decays of the nucleosynthesis products produced in novae
is such an approach; these decays in general emit positrons, whose
annihilation with electrons produces a line at 511 keV.  An obvious
advantage of this approach is the very small absorption of $\gamma$-rays
in the Galaxy.  We will also see that problems of uneven coverage and
sensitivity are minimal.  These advantages are realised when the 
$\gamma$-ray detector TGRS, on board the {\em WIND\/} mission, is used 
(\S 2.1).

In Paper I we determined
that TGRS does indeed have the capability to perform a sky survey for
classical novae.  The target of Paper I was to detect the positron 
annihilation line in five known novae; although none was detected, the 
viability of such a method was established.  The key to the method (see
\S 2 below) is that the line arises in nova material expanding towards 
the observer, and is therefore broadened and blueshifted (Leising \&
Clayton 1987).  Its peak is therefore shifted away from a strong background
line at exactly 511 keV, which arises in the instrument itself from decays
of unstable nuclei produced by cosmic ray spallation.

In the next section we give a brief description of the detector and data,
and of our analysis.  None of these is substantially different from that
of Paper I, where the reader may find a more detailed description. 

\section{Observations and Analysis}

\subsection{Spacecraft and Instrument}

The TGRS experiment is very well
suited to a search for the 511 keV line, for several reasons.  First,
it is located on board the {\em WIND\/} spacecraft whose orbit is so 
elliptical that it has spent virtually all of its mission since 1994 
November in interplanetary space, where the $\gamma$-ray background level 
is relatively low.  Second, these backgrounds are not only low but very
stable over time.  Third, TGRS is attached to the south-facing surface
of the rotating cylindrical {\em WIND\/} body, which points permanently
toward the south ecliptic pole.  The detector is unshielded, and TGRS
therefore has an unobstructed view of the entire southern ecliptic
hemisphere.  Taken together, these three facts make possible a continuous
and complete survey of the southern sky.  Fourth, and most importantly, 
the TGRS Ge detector has
sufficient spectral resolution to detect a 511 keV line which is slightly
Doppler-shifted away from the background 511 keV line mentioned in \S 1.  
The Doppler blueshift in the nova line, for the epochs $<$12 hr 
which we consider, is predicted to be 2--5 keV (Hernanz 1999,
Kudryashov 2000), which compares with the TGRS energy resolution at 511 keV 
of 3--4 keV FWHM (Harris et al. 1998 and Paper I).

The TGRS detector is a radiatively cooled 35 cm$^{2}$ n-type Ge crystal
sensitive to energies between 20 keV--8 MeV.  Since the launch of
{\em WIND\/} in 1994 November, TGRS has accumulated count
rates continuously in this energy range.  The few gaps in the data
stream are due either to perigee passes, which are rare (lasting 
$\sim 1$ d at several month intervals) thanks to {\em WIND\/}'s
very eccentric orbit, or to memory readouts following solar flare or
$\gamma$-ray burst triggers, which may last for $\sim 2$
hr.  The data were binned 
in 1 keV energy bins during 24 min intervals.  

We searched in data covering a period of nearly three years, from 1995 
January to 1997 October.  In the fall of 1997 the performance of the
detector began to degrade seriously, and the energy resolution became
too coarse to resolve the 511 keV background and nova lines.  This
degradation is believed to result from crystal defects induced by
accumulated cosmic ray impacts, which trap semiconductor holes and
reverse the impurity charge status.  A region of the crystal thus
becomes undepleted and the effective area is reduced (Kurczynski et
al. 1999).  We terminated our search of the data when the photopeak
effective area at 511 keV fell below an estimated 80\% of its 
original value.  The total live time accumulated was about
$7.7 \times 10^{7}$ s, which was nominally 88\% of the entire
interval.  In fact, the distribution of live times among the 6 hr
intervals was such that 41\% of all intervals had the full 6 hr of
live time, and almost 99\% of intervals contained some
live time.

\subsection{Analysis}

Our analysis procedure relies heavily on the most recently 
theoretically-predicted properties of the 511 keV line (Hernanz et
al. 1999, Kudryashov 2000), mainly its light-curve, energy and
shape.  The timescale over which the background spectra described
above are summed is set by the predicted $\gamma$-ray light-curve from the
"thermonuclear flash" which powers a nova.  In this process a degenerate 
accreted H layer on
the surface of a CO or ONe white dwarf ignites proton
capture reactions involving both accreted material and some material
dredged up from the interior of the white dwarf.  The
timescale for this process is set by the $\beta$-decay timescales
of the unstable nucleosynthesis products of rapid proton capture on
C, O and Ne.  These unstable species fall into two groups, one having 
very rapid decays ($\sim$ minutes: e.g. $^{13}$N, $^{14,15}$O,
$^{17}$F) and the more slow-decaying $^{18}$F ($\tau_{1/2} =  110$ min).
The light-curve results from the convolution of these decays
with the reduction of opacity to 511 keV $\gamma$-rays due to envelope
expansion; it thus tends to be double-peaked at values $\sim$10--100
s and $\sim$3--6 hr (G\'{o}mez-Gomar et al. 1998), with 
significant emission lasting for $\sim 12$ hr
(Hernanz et al. 1999).  The 10--100 s peak is ultimately due to the
decay of the very short-lived group of isotopes, and is thus especially
prominent in the CO nova light-curve (though these isotopes are essential
to the energetics of both classes).  The 3--6 hr peak reflects the survival of
slower-decaying $^{18}$F in both classes (G\'{o}mez-Gomar et al. 1998).

With these timescales in mind, we summed the 24-min background
spectra into 6 hr intervals \footnote{ Not only does this
accumulation time contain most of the predicted line fluence, but
detection of the line in the following 6 hr period would provide an
independent confirmation of a detection.  However, we found in
Paper I that detection of the line after 12 hr is hindered not only
by low flux, but also by worsened blending with the background 511
keV line.  The details of how measurements in the 6--12 hr interval
may be combined with those in the 0--6 hr interval are given in Paper
I.}  The 4005 resulting 6-hr spectra were fit by a model (described
in Paper I) containing the strong background 511 keV line at rest,
and a broadened blueshifted nova line.  The energies of the nova line
were fixed at the predicted values (516 keV after 6 hr, dropping to
513 keV after 12 hr: G\'{o}mez-Gomar et al. 1998, Hernanz et al.
1999, Kudryashov 2000).  The widths were taken to be 8 keV FWHM and
the shapes to be Gaussian, as in Paper I; the shapes are poorly
documented in published models, but the approximation is probably
reasonable at an epoch of a few hours (Leising \& Clayton 1987).  
Instrumental broadenings of these lines and of the background 511 keV line 
were very small during 1995--1997 (Harris et al. 1998).  Although our
analysis is somewhat sensitive to the departure of the actual line
parameters from these predictions, we believe that it should be
adequate to detect lines in the parameter range appropriate for fast
novae.  For example, we estimate that lines with energies in the range
513--522 keV are detected with $\ge 50$\% of true amplitude, corresponding
to expansion velocities 1200--6500 km s$^{-1}$ which bracket the range 
observed in fast novae (Warner 1995).

The 4005 count spectra were fit to the above model (plus an underlying
constant term) and the line amplitudes were divided by the photopeak
effective area at 511 keV.  This photopeak efficiency was determined from
Monte Carlo simulations as a function of energy and zenith angle (Seifert
et al. 1997), taking into account the effects of hole-trapping in the detector
(\S 2.1); we found that the efficiency remained extremely stable until
the fall of 1997, whereupon it rapidly fell to 80\% of its
value at launch.  The effective area is a slowly varying function of the
zenith angle of the source.  To calculate the average effective area, we 
assumed the Galactic distribution of the synthetic
population of several thousand novae computed by Hatano et al. (1997),
for the southern part of which the mean TGRS zenith angle is $60 \deg$,
corresponding to an effective area 13.6 cm$^{2}$.  
The fits were performed by the standard method of varying the model
parameters to minimize the quantity $\chi^{2}$, with errors on the
parameters computed from the parameter range where $\chi^{2}$
exceeded the minimum by +1 (Paper I).

With a sufficiently large sample of spectra, there is a probability
that a fitted line of any given amplitude may be produced by chance.
We therefore imposed a rather high value of significance as the
threshold above which a detection would be established.  If the
significances are normally distributed (see \S 3 below) then
our sample size of 4005 spectra implies
that a threshold level of $4.6 \sigma$ yields a probability
$< 1$\% of a single false detection by chance (Abramowitz \& Stegun 1964).

\section{Results}

A typical fit to a 6 hr spectrum is shown in Figure 1 (there is a
more detailed discussion in Paper I).  The fits are generally
acceptable, with values of $\chi^{2}$/d.o.f. close to 1.  The
amplitudes of the nova lines are significantly positive in all
cases (see below).  This arises from a significant departure
of the blue wing of the 511 keV background line from the Gaussian
shape assumed in the fits, whose origin is unclear (Paper I).

The full series of measurements for a nova line of FWHM 8 keV and
blueshift 5 keV, (parameters corresponding to typical predicted values 
after 6 hr: Hernanz et al. 1999 and Paper I)
is shown in Figure 2.  It can be seen that the systematic
positive offset mentioned above was extremely stable throughout
the mission; there are very weak linear trends on $\sim$year 
timescales which are almost invisible in Fig. 2.  We subtracted
this quasi-constant systematic value from all nova line
measurements.

A very similar time series was obtained for a nova line at
position predicted for 12 hr after the explosion (513 keV), except
that the error bars were very much larger (see Paper I, \S 3.4).
Each fitted 6 hr line was combined with the following 12 hr fit in
the proportions suggested by the light curve of Hernanz et al. (1999).  
The results closely resembled those of Fig. 2 after
subtraction of the quasi-constant systematic, since the 12 hr lines
contributed little on account of their large error bars.

It is also clear from inspection of Fig. 2 that there are no highly
significant line amplitudes lying above the mean.  We further show
in Figure 3 that the distribution of significant deviations from the
mean is very close to normal.  The variability in the error bars 
comes almost entirely from the variability in live times, which is
small (\S 2.1).  There is therefore a well-defined mean $1 \sigma$ 
error of $8.2 \times 10^{-4}$ photon cm$^{-2}$ s$^{-1}$ (compare
results of Paper I for zenith angle $60 \deg$).  The $4.6 \sigma$
threshold based on this average error is shown by a dotted line in
Fig. 2.  The only points lying above this line are a few 6 hr 
periods with low live time and large errors.  We therefore 
conclude that no previously-undetected novae were discovered by TGRS 
during 1995--1997, in an almost unbiased search covering a live time
of $7.7 \times 10^{7}$ s.

\section{Discussion}

\subsection{Limit on the Galactic nova rate}

Recent developments in the theory of nucleosynthesis in classical novae
(Hernanz et al. 1999) have been discouraging for our purpose of a
positron annihilation $\gamma$-ray search, since new measurements of
nuclear reaction rates have led to much lower predictions of the flux
in this line after 6 and 12 hr.  The discussion in Paper I of the
capability of constraining the global Galactic nova rate using our present
results was therefore over-optimistic.  Nevertheless, we will
discuss the application of our method in general terms, so that even 
though important constraints cannot now be derived, it may be useful
for more sensitive future experiments (e.g. {\em INTEGRAL\/}) or for
more optimistic theoretical predictions.

A formal expression for the number of novae detectable by TGRS is
\begin{eqnarray}
N_{obs} & = & R_{gal} ~T_{tot} ~\int_{0}^{1.4 M_{\sun}}
~\Phi(M) \int ~f(\phi > \phi_{min}) ~w(\phi > \phi_{min}) ~d\phi_{min} ~dM
\end{eqnarray}
where $R_{gal}$ is the Galactic nova rate; $\phi_{min}$ is 
a given (time varying) threshold flux for detection by 
TGRS; $f(\phi > \phi_{min})$ is the fraction of the mass of the Galaxy 
within TGRS's detection radius $r_{d}$ and $r_{d}  = 
\sqrt{\phi_{pred}(M)/\phi_{min}}$; $T_{tot}$ is the total TGRS
live time; $w(\phi > \phi_{min})$ is the fraction of TGRS live time for
which $\phi > \phi_{min}$; $\Phi$ is the distribution of white dwarf masses in
classical novae; and $\phi_{pred}(M)$ is the predicted 511 keV line
flux at 1 kpc for mass $M$.  

The white dwarf mass distribution in novae, $\Phi(M)$, is very poorly
known.  Whereas field white dwarf masses appear to peak at $\simeq
0.6 M_{\sun}$ and to decline in number for higher masses up to the
Chandrasekhar limit $1.4 M_{\sun}$ (Warner 1990), the mass distribution
in nova systems must be weighted towards higher masses.  This is because
the thermonuclear runaway occurs when the basal pressure of the material
accreted onto the white dwarf exceeds some critical value.  The critical
pressure is proportional to the -4 power of the white dwarf radius, to
the white dwarf mass, and
to the accreted mass.  Since white dwarf radii decrease with increasing
white dwarf mass, the accreted mass necessary to reach critical pressure
is a strongly decreasing function of white dwarf mass.  If the accretion
rate from the secondary star is roughly independent of white dwarf mass,
it follows that explosions on more massive white dwarfs will recur after
much shorter intervals (Gehrz et al. 1998).
There have not been reliable measurements of this effect, although
theory indicates that the ratio of ONe:CO novae of 1:2 is compatible
with a distribution peaking at about $1.2 M_{\sun}$ (Truran \& Livio
1986).  Further, the mass ranges corresponding to the CO and ONe
compositions are poorly known and may well overlap (Livio
\& Truran 1994).

Theoretical predictions of 511 keV line emission are only
available for a few values of $M$.  In Table 2 we show the parameters
of the most recent models suitable for use in Eq. (4) (Hernanz et al.
1999).  Earlier models suggest that emission from lower-mass CO
white dwarf events is considerably less (G\'{o}mez-Gomar et al. 1998).
In view of the remarks above about the ONe:CO ratio, we will make the
crude assumption that the ratio of "low mass" CO objects to "high mass" CO
objects to ONe objects is 1:1:1, where "high mass" CO objects have the
properties given in Table 2 and "low mass" CO objects are assumed to
produce no 511 keV line emission at all.  This eliminates the integral
over {\em M\/} in Eq. (4).

The remaining integral in Eq. (4) can be approximated by the value
of the integrand when $\phi_{min}$ has its mean value --- this follows
from our result in \S\S 2.1, 3 that the variation of live times in 
our sample (and therefore of the errors in Fig. 2) is very small.
For a given model in Table 2, therefore, taking 
$\phi_{min} = 4.6 \times 8.2 \times 10^{-4} 
= 3.8 \times 10^{-3}$ photon cm$^{-2}$ s$^{-1}$, the problem is
reduced to the computation of the fraction $f$ of Galactic mass which
lies within the radius 
$r_{d} = \sqrt{\phi_{pred}/3.8 \times 10^{-3}}$ kpc.  

As an example, let us consider the Hernanz et al. (1999)
model of a $1.15 M_{\sun}$ CO nova from Table 2.  Here $\phi_{pred} =
3.1 \times 10^{-3}$ photon cm$^{-2}$ s$^{-1}$, so that $r_{d}
= 0.9$ kpc.  Within this value of $r_{d}$ we determined that 0.61\%
of the Galaxy's mass resides, according to
the widely used Bahcall-Soneira model of the Galaxy (Bahcall \&
Soneira 1984): one-half of this value (i.e. the southern hemisphere)
gives $f = 0.00305$.  For the $1.15 M_{\sun}$ CO model, Eq. (1) then reduces
to $R_{gal} =\frac{N_{obs}}{T_{tot} ~f}$.  Our upper limit for
$N_{obs}$ is $<1$, with 63\% probability (for Poisson-distributed
events: Gehrels 1986), and the live
time is $7.7 \times 10^{7}$ s (\S 2.1), giving us a 63\% upper
limit on the rate of "high-mass" CO novae of $R_{gal} <134$ yr$^{-1}$.  
This is quite close to the exact value 123 yr$^{-1}$ obtained by
explicitly integrating Eq. (4) over $\phi_{min}$ (Table 2).

In the same way we obtain an upper limit of 238 yr$^{-1}$ 
on the rate of novae
occurring on ONe white dwarfs from Hernanz et al.'s (1999) prediction.  
Our best result comes from the CO model in Table 2, which we
have assumed to be
one-third of the total, from which we derive a global Galactic nova rate of 
$< 369$ yr$^{-1}$.  

Uncertainties in this value clearly arise from uncertainties in the nova
models, in the fraction of white dwarfs in nova systems of each type,
in the Bahcall-Soneira model, and in the possibility of
distinct spheroid and
disk nova populations having differing rates, since our typical 
detection radius $< 1$ kpc includes almost none of the
Bahcall-Soneira spheroid.  Since our results do not significantly
constrain previous measurements of the nova rate (Table 1), we do not
make estimates of these errors, which will require attention from
other, more sensitive experiments (see next section).

\subsection{Implications for nova detection by other instruments}

An attempt has been made, using the BATSE instrument on the {\em Compton\/}
Observatory, to detect 511 keV line emission from a recent nearby nova
(V382 Vel) by a similar method to that used here and in Paper I  
(Hernanz et al. 2000).  The advantage of 
observing with BATSE over TGRS is its much larger effective area.  Its 
disadvantages are much poorer energy resolution with a NaI spectrometer,
and a background varying on very short ($<< 90$ min) timescales.  The
sensitivities achieved are comparable to those obtained here.

Degradation of the Ge detector (\S 2.1) prevented TGRS 
from achieving comparable sensitivity on V382 Vel (Harris et al. 2000),
so future efforts in this field will rely on BATSE and on
the {\em INTEGRAL\/} mission, which is scheduled for launch in 2001 September 
carrying a Ge spectrometer (SPI) with resolution comparable to TGRS but a much
larger effective area.  Hernanz et al. (1999) estimated 
that SPI could detect the model novae of Table 2 out to $\sim 3$ kpc.
However they also pointed out that the short duration of the 511 keV line 
emission would make it difficult for {\em INTEGRAL\/} to slew to a candidate
event.  Thus the detection rate would be limited to novae within the
SPI field of view, which is $\sim 25 \deg$ FWHM.

The search method which we have used, i.e. an {\em ex post facto\/} search
in background spectra, ought to be perfectly feasible with {\em INTEGRAL\/}.  
The chief requirements for this method are very high energy resolution and a
sufficiently low and stable background.  While the SPI detector has 
excellent resolution, the background level in it has not yet been
rigorously computed.  Nevertheless, qualitative arguments suggest that
the background will be no worse than that in TGRS.  Like {\em WIND\/},
{\em INTEGRAL\/} will be in a high-altitude elliptical orbit which
avoids extensive exposure to Earth's trapped radiation belts and to
albedo $\gamma$-rays from Earth's atmosphere.  The main disadvantages of
{\em INTEGRAL\/} for nova detection are the small SPI field of view and
the planned observing strategy which cuts down the amount of time spent
pointing towards the main concentration of novae near the Galactic center.

We can make use of the planned program of
{INTEGRAL\/} observations of the central Galactic radian in the first
year of operation (Winkler et al. 1999) to estimate the rate at which
novae might be detected in the SPI data.  As previously, we assume that
novae follow the Bahcall \& Soneira (1984) Galactic distribution. 
The planned first-year {\em INTEGRAL\/} observations may be approximated by
a $31 \deg \times 11 \deg$ grid with $2 \deg$ spacing between 
$-30 \deg \ge l \ge 30 \deg$ and $-10 \deg \ge b \ge 10 \deg$, the
exposure to each point being 1180 s per pass, with 12 passes per
year covering the whole grid.  Thus the live time for the whole grid 
is 0.153 yr.  From the Bahcall-Soneira model, the pointing
geometry, and the SPI aperture $\sim 25 \deg$ we calculated that the
{\em INTEGRAL\/} detection radius $\sim 3$ kpc intercepts
$\sim$0.75\% of the Galactic nova distribution.  The live time
0.153 yr is then multiplied by a typical Galactic nova rate $\sim 50$ yr$^{-1}$,
(of which 2/3 are practically detectable, as assumed in \S 4.1),
and by the intercepted fraction, to 
imply that {\em INTEGRAL\/} ought to detect 0.04 novae yr$^{-1}$. 
Unless theoretical estimates of the 511 keV line flux turn out to be
considerably larger, the prospects for such a detection appear to be
small.  The same conclusion probably applies to a different method
of detecting 511 keV line emission indirectly, by observing the 170--470
keV continuum produced by Compton scattering in the nova envelope using
SPI's large-area CsI shield (Jean et al. 1999).

\acknowledgments

We are grateful to M. Hernanz and A. Kudryahov for helpful discussions and 
for providing pre-publication results, and to J. Jordi (the referee) for
constructive comments.  Peter Kurczynski (University
of Maryland) helped in assessing the instrument performance.
Theresa Sheets (LHEA) and Sandhia Bansal (HSTX)
assisted with the analysis software.

\clearpage

\begin{figure}

\caption{Characteristic TGRS background count spectrum at energies around
511 keV, obtained during the
interval 3 June 12h--18h UT.  (a) Fit to the blue wing of the TGRS
background line at 511 keV with
the power law continuum subtracted.  Apart from the power law, the
components of the fit are a line with the 
width and position of the nova 6-hr line
(dot-dashed line) and a Gaussian line fitting the blue wing of the
511 keV background line (dashed line).  The total model spectrum
is the full line.  (b)  Expansion of Fig. 1a showing the significance of the
fitted nova 6-hr line (dot-dashed line of amplitude $5.2 \pm 0.1 \times
10^{-3}$ photon cm$^{-2}$ s$^{-1}$; other symbols as in Fig. 3a).
Also shown is the theoretically expected level of the nova 6-hr line for
a nova at 500 pc (Hernanz et al. 1999, CO model).}

\caption{(a) Measured fluxes in the nova 6-hr line during the entire
interval 1995 January -- 1997 October.  Dotted
line --- mean $4.6 \sigma$ upper limit, above which candidate line
detections would lie.}

\caption{Distribution of significances of the flux measurements in
Fig. 2 (with constant sytematic subtracted).  Dashed 
line --- Gaussian distribution of unit width and normalization.}

\end{figure}

\clearpage

\begin{table*}
\begin{center}
\begin{tabular}{lcc}
\tableline
Method & Rate yr$^{-1}$ & Reference \\
\tableline
M31, M33, LMC comparison & $24^{+26}_{-9}$ & (1) \\
Extrapolate from known nearby novae & $\sim 34$ & (2) \\ 
Correct incompleteness \& extinction & $73 \pm 24$ & (3) \\
Correct incompleteness \& extinction & $35 \pm 11$ & (4) \\
Correct incompleteness \& extinction & 260 & (5) \\
Monte Carlo simulation & $41 \pm 20$ & (6) \\
M31 comparison & 46 & (7) \\
Extrapolate luminosity function & $\sim 175$ & (8) \\ 
External galaxies comparison & 11--46 & (9) \\
M31 comparison & $<13$ & (10) \\

\tableline

\end{tabular}
\end{center}

{\em REFERENCES\/} \\
1. DellaValle \& Livio 1994.\\
2. Warner 1989.\\
3. Liller \& Mayer 1987.\\
4. Shafter 1997.\\
5. Sharov 1972.\\
6. Hatano et al. 1997.\\
7. Higdon \& Fowler 1987.\\
8. Allen 1954.\\ 
9. Ciardullo et al. 1990. \\
10. Van den Bergh 1988. \\
11. Present work. \\

\caption{Estimates of the global Galactic nova rate}

\end{table*}

\clearpage

\begin{table*}
\begin{center}
\begin{tabular}{lcc}
\tableline
Model & CO & ONe\\
\tableline
White dwarf mass, $M_{\sun}$ & 1.25 & 1.15 \\
Line flux at 1 kpc,\tablenotemark{a} & $3.1 \times 10^{-3}$ & 
$1.8 \times 10^{-3}$ \\
~~photon cm$^{-2}$ s$^{-1}$ & & \\
Detection radius $r_{d}$, kpc & 0.9 & 0.7 \\
Galactic rate, yr$^{-1}$ & $<123$ & $<238$ \\
\tableline

\end{tabular}
\end{center}

\tablenotetext{a}{ Estimated average flux during the first 6 hr of the
explosion from the light-curve of Hernanz et al. (1999).}

\caption{Galactic nova rate for recent models of 511 keV line emission}

\end{table*}

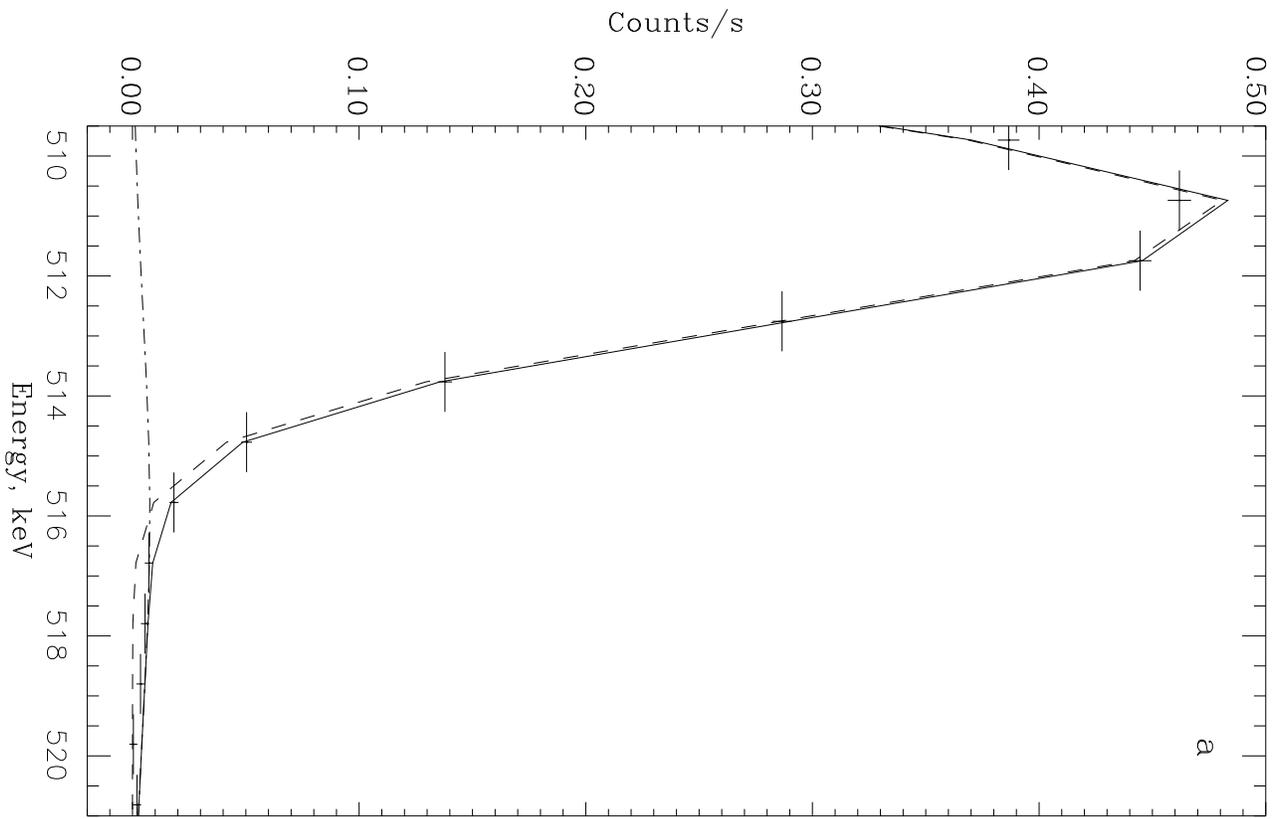
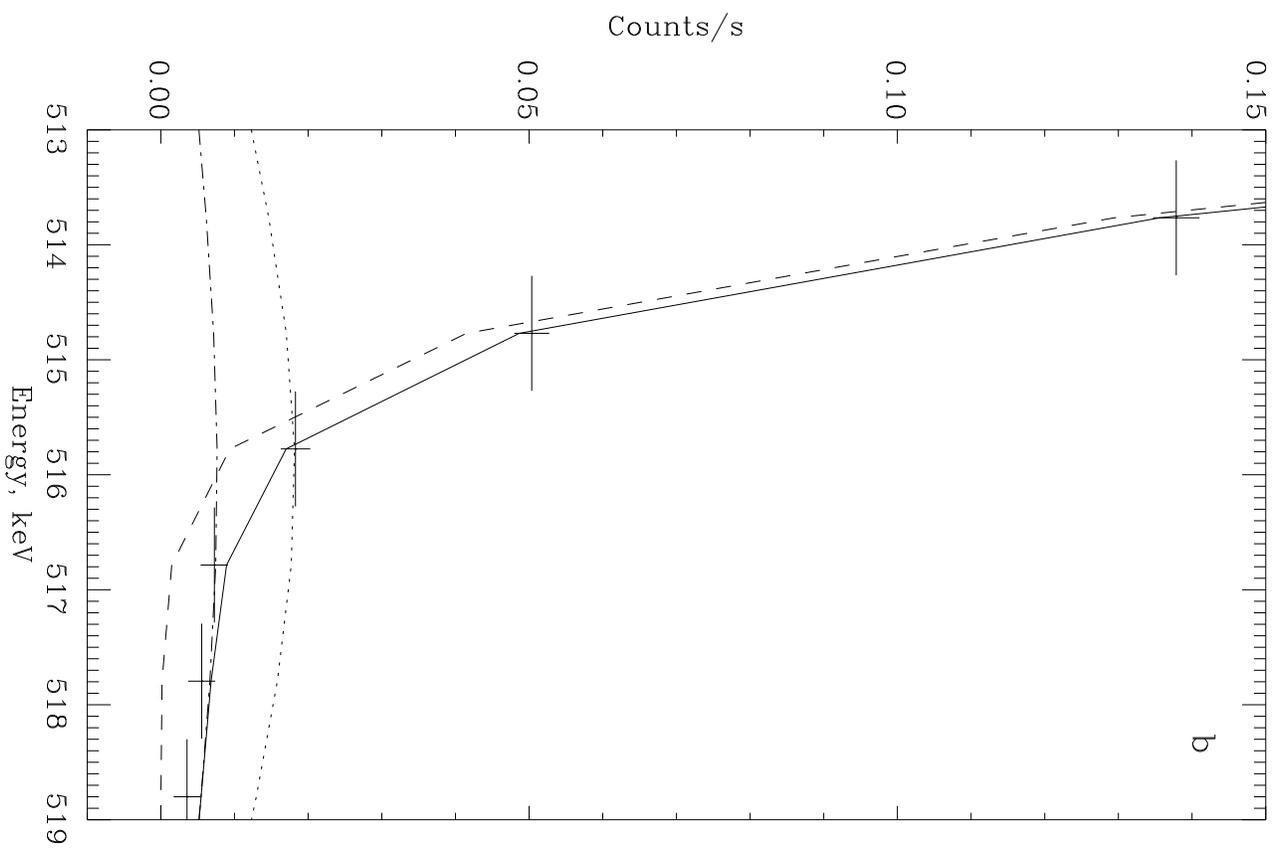
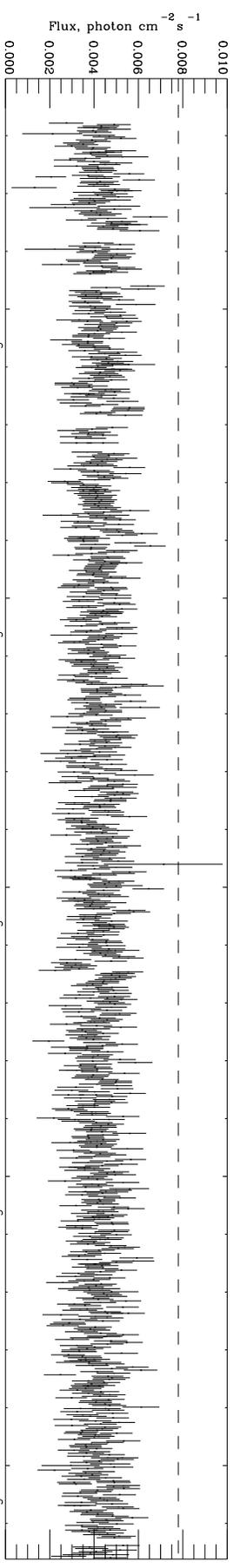
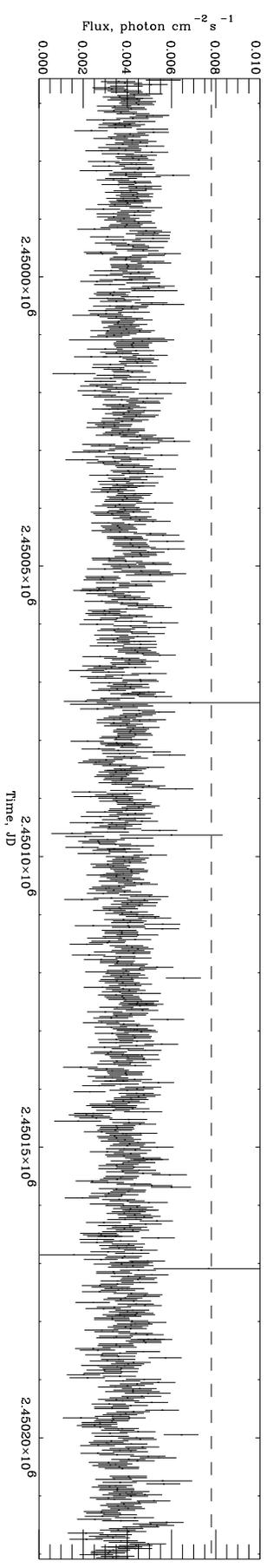
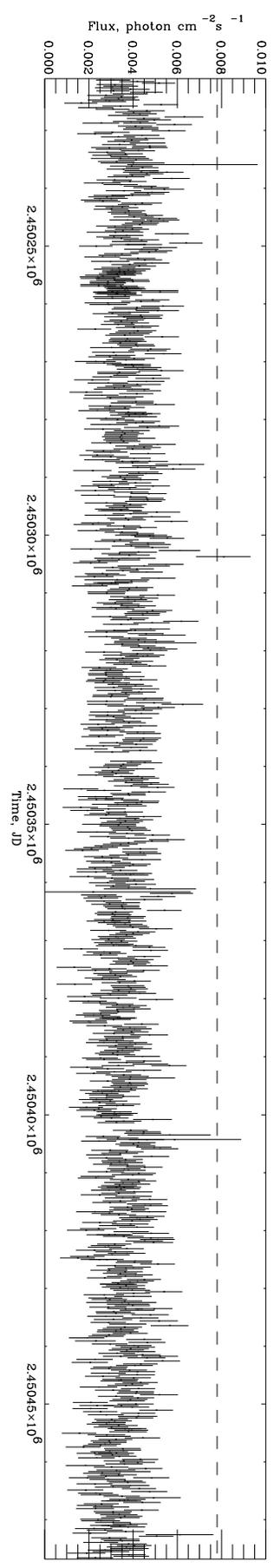
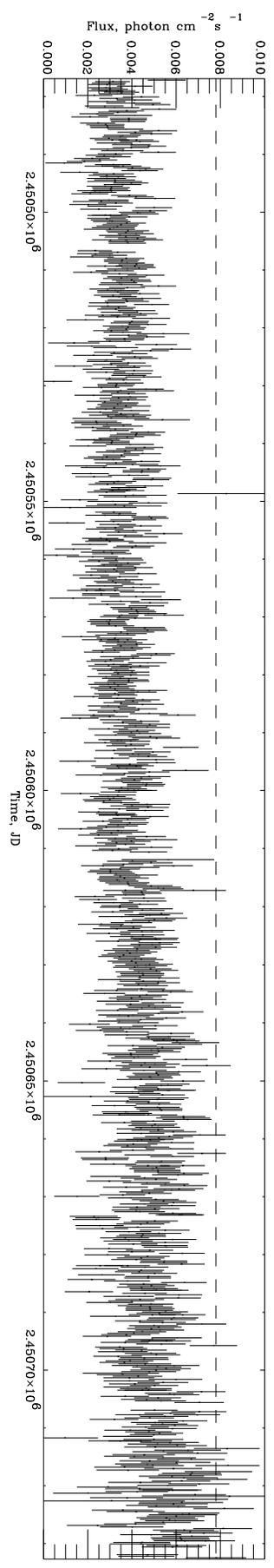
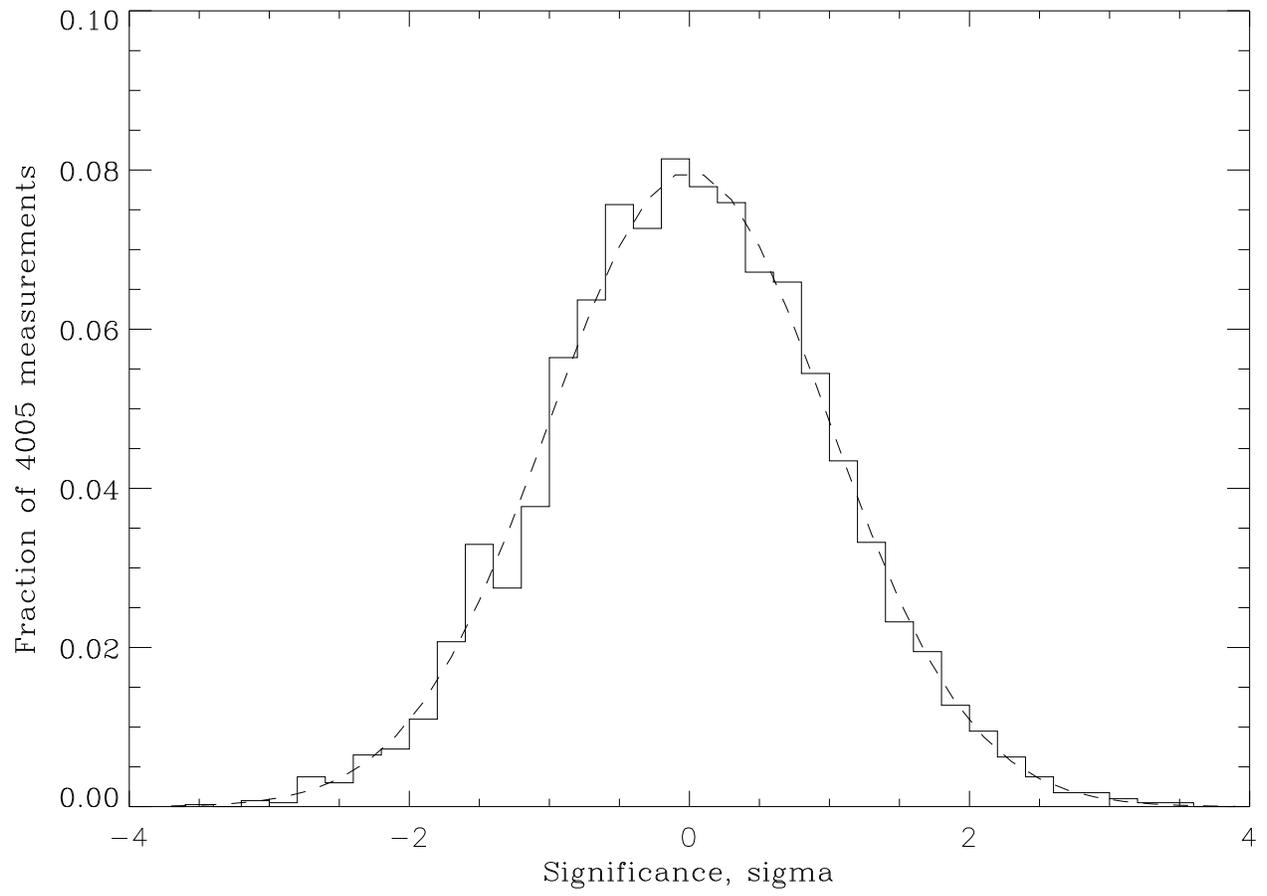
\end{document}